\documentclass[amssymb,twocolumn,eqsecnum,aps]{revtex4}
\usepackage{amsmath}
\begin{document}
%\centerline{File: manuscripts/part2.tex}
\title{Electrodynamics in a Filled Minkowski Spacetime with
Application to Classical Continuum Electrodynamics}
\author{Michael E. Crenshaw}
\affiliation{AMSRD-AMR-WS-ST, USA RDECOM, Aviation and Missile RDEC,
Redstone Arsenal, AL 35898, USA}
\date{\today}
\begin{abstract}
Minkowski spacetime is a convenient setting for the study of the
relativistic dynamics of particles and fields in the vacuum.
In order to study events that occur in a dielectric or other linear
medium, we adopt the familiar continuum assumption of a linear,
isotropic, homogeneous, transparent medium of refractive index $n$
filling all space and seek the principle of relativity that applies in
the filled spacetime.
Applying the Einstein postulates with $c/n$ as the speed of light,
we show how the effective signal velocity results in a scaling of the
proper time by the refractive index and examine the consequences
for D'Alembert's principle, the Lagrange equations, and the canonical
momentum field.
The principles of dynamics in the filled spacetime are then applied to
the electromagnetic Lagrangian and we derive equations of motion
that are invariant with respect to a material Lorentz transformation.
The new representation of the dynamics of macroscopic fields is
shown to be consistent with the equal-time commutation relation
for quantized macroscopic fields, quantum--classical correspondence,
the principle of superposition, and electromagnetic boundary conditions.
\end{abstract}
\maketitle
\vskip 3.0cm
\par
\section{Introduction}
\par
The basic tenant of classical continuum electrodynamics is that the
electrodynamic properties of a material can be characterized by
macroscopic parameters.
While linear media are not homogeneous on a microscopic scale, most
optical phenomena take place on a characteristic scale of many unit
cells for which the macroscopic model of a linear medium is a useful
abstraction. 
For the purpose of electrodynamics, a transparent linear medium can be
thought of as a collection of atoms or molecules, behaving like simple 
harmonic oscillators, embedded in the vacuum of Minkowski spacetime.
On length scales smaller than the spacing between dipoles, a light
signal travels away from an event at the velocity $c$, defining the null
surface of a light cone as $|\Delta {\bf x}|^2=(c\Delta t)^2$ on a
four-dimensional Minkowski map \cite{BIRindler,BISchwarz,BICarroll}.
At substantially longer length scales, the light signal from an event
travels at an effective velocity $c/n$, defining a macroscopic
refractive index $n$.
Then the null-surface is an effective light cone that is described by
$|\Delta {\bf x}|^2=(c\Delta t/n)^2$ \cite{BIFinn}.
If we are not interested in the behavior of events on the atomic scale,
then we can apply the macroscopic model of a linear medium as a
continuum with macroscopic parameters.
In this continuum limit, electrodynamics takes place on the background
of an effective Minkowski spacetime with a timelike coordinate
$x_0=ct/n$.
\par
The theory of relativity defines transformations between different
inertial reference frames moving at constant velocities
 \cite{BIRindler,BISchwarz,BICarroll}.
Einstein showed that the microscopic Maxwell equations are invariant
under transformations of the vacuum Lorentz group.
Like the microscopic Maxwell equations for free-space, the macroscopic
Maxwell equations of continuum electrodynamics need to be invariant
under some transformation \cite{BIFinn1}.
We consider a linear isotropic homogeneous medium of refractive index
$n$ filling all space and seek the principle of relativity that applies
in this spacetime in which light travels at speed $c/n$.
Adopting the Einstein postulates with $c/n$ as the speed of light in
the filled spacetime results in a material Lorentz
transformation \cite{BIFinn1} and a scaling of the
proper time interval by the refractive index \cite{BIquantclass}.
Here, we derive the field dynamics under these conditions and examine
the consequences for continuum electrodynamics.
\par
There are a number of issues that contribute to make this undertaking
necessary.
i) Macroscopic electromagnetic fields in linear materials, when
quantized using current procedures 
\cite{BIMacro1,BIMacro3,BIMacro4,BIMacro5},
violate quantum--classical
correspondence in relation to the electromagnetic boundary
conditions \cite{BIquantclass}.
A re-derivation of D'Alembert's principle and the Lagrange equations of
motion for discrete particles in a dielectric-filled Minkowski spacetime
was sufficient to repair the violation of the correspondence principle
\cite{BIquantclass}.
The appearance of the refractive index in the time-like coordinate
of filled spacetime and in the proper time that modifies the dynamics
of discrete particles will affect field dynamics, as well.
ii) The definition of the linear refractive index in terms of the linear
electric and magnetic susceptibilities violates the principle of
superposition.
Ward, Nelson, and Webb \cite{BIWNW} used the Feynman \cite{BIFeynLect}
definition of the refractive index in terms of the effect of a
refractive medium on a transmitted field to derive the effective index
$n=1+\chi_e+\chi_m$ of a magnetodielectric material in terms of the
the electric susceptibility $\chi_e$
and the magnetic susceptibility $\chi_m$.
If either the electric or magnetic susceptibility is small
compared to unity, which is usually the case,
then $n^2\approx(1+\chi_e)(1+\chi_m) =\varepsilon\mu$ in terms of the
permittivity $\varepsilon$ and permeability $\mu$.
However, the approximation is not valid if the electric and 
magnetic susceptibilities are not small compared to the unit vacuum
susceptibility $\chi_0=1$ and we make the case for
accepting the derived result $n=\chi_0+\chi_e+\chi_m$,
which obeys superposition.
The case for the principle of superposition has also been argued from
quantum electrodynamics based on the linearity of quantum
mechanics \cite{BIsuperposition}.
%iii) A century on, the Abraham--Minkowski controversy for the
%momentum of the electromagnetic field in a linear material still has
%no single final resolution.
%The persistance of a not-too-subtle ambiguity in a well-established
%physical theory indicates that the issue cannot be resolved within the
%existing formalism.
%The presence of the refractive index in the timelike coordinate of
%filled spacetime and in the proper time has been shown to affect 
%the momentum of particles in a region of space in which light travels
%at $c/n$.
%This is also true for fields necessitating a
%re-examination of the issue of momentum of the electromagnetic field in
%a linear medium.
iii) The transformation properties of electromagnetic fields in linear
media are a consequence of the continuum assumption, which is requisite
for continuum electrodynamics.
In a linear isotropic homogeneous continuous medium, the equations of
motion must therefore be constructed for invariance under the material
Lorentz group, rather than the vacuum Lorentz group \cite{BIFinn1}.
\par
In this article, the relativity of events in a space-filling linear
medium is used in the derivation of Hamilton's equations of motion for
fields.
The Lagrangian dynamics of discrete particles in a filled Minkowski
spacetime that was presented in Ref.\ \cite{BIquantclass} is reviewed.
The field theory in a linear medium is then developed as a
generalization of the discrete theory.
The canonical momentum field is found to depend on the presence of
the refractive index in the timelike coordinate of the filled
Minkowski spacetime through differentiation with respect to the proper
time.
Using the field theory, we derive macroscopic equations of motion
\begin{subequations}
\label{EQb1.01}
\begin{equation}
\nabla\times{\bf B}=- \frac{\partial{\bf \Pi}}{\partial x_0}
\label{EQb1.01a}
\end{equation}
\begin{equation}
\nabla\times{\bf \Pi}= \frac{\partial{\bf B}}{\partial x_0}
\label{EQb1.01b}
\end{equation}
\end{subequations}
for the electromagnetic field in a linear medium, where
${\bf B}=\nabla\times{\bf A}$ is the magnetic field,
${\bf \Pi}= \partial{\bf A}/\partial x_0$ is the conjugate momentum
field, and $x_0=ct/n$ is the timelike coordinate of filled Minkowski
spacetime.
Before dismissing this derived result {\it proctor hoc} with respect to
the macroscopic Maxwell equations, it should be noted that the use of
the timelike coordinate of filled Minkowski spacetime $x_0 = ct/n$,
rather than the time $t$ or the timelike coordinate $x_0 = ct$, of
vacuum Minkowski spacetime is mandated by Lorentz invariance under the
material Lorentz group.
As shown here, the canonical momentum field in the macroscopic Maxwell
equations leads to a violation of quantum--classical correspondence
between the Fresnel boundary conditions and the equal-time commutation
relation for quantized fields.
Using the Feynman model of the refractive index, we show that the
vacuum, electric, and magnetic susceptibilities obey the principle of
superposition in the new dynamical theory.
We also show that the application of Stokes' theorem and conservation
of energy to the macroscopic equations of motion (\ref{EQb1.01})
can be used to derive the Fresnel boundary conditions.
Then the principle result of this work is the set of dynamical equations
(\ref{EQb1.01}) for electromagnetic fields in linear media that are
consistent with Lorentz transformations and special relativity,
quantum--classical correspondence and equal-time commutation relations,
the superposition principle, and electromagnetic boundary conditions.
\par
\section{Relativistic Particle Dynamics in a Filled Spacetime}
\par
A distribution of particles is, in the continuum approximation, regarded
as a continuous medium and a property of the particles can be
represented by a property density that is a continuous function of the
spatial and temporal coordinates.
In continuum electrodynamics, a linear medium is a uniform region of
space in which light travels from a source to an observer at a constant
speed of $c/n$.
In Ref.\ \cite{BIquantclass}, we considered space to be entirely filled
with an isotropic homogeneous continuous linear medium and derived the
characteristics of spacetime and relativity for the case in which the
speed of light is $c/n$.
The purpose of the current work is to generalize the dynamics of 
discrete particles to derive the dynamics of electromagnetic fields in
a filled Minkowski spacetime.
\par
Consider two inertial reference frames, $S(t,x,y,z)$ and
$S^{\prime}(t^{\prime},x^{\prime},y^{\prime},z^{\prime})$, in a
standard configuration in which $S^{\prime}$ translates at a constant
velocity $v$ in the direction of the positive $x$ axis and the origins
of the two systems coincide at time $t=t^{\prime}=0$.
If a light pulse is emitted from the common origin at time $t=0$, then
\begin{equation}
x^2+y^2+z^2-\left ( \frac{ct}{n} \right ) ^2=0 
\label{EQb2.01}
\end{equation}
describes wavefronts in the $S$ system and
\begin{equation}
(x^{\prime})^2+ (y^{\prime})^2+ (z^{\prime})^2
-\left (\frac{ct^{\prime}}{n} \right )^2=0 
\label{EQb2.02}
\end{equation}
describes wavefronts in the $S^{\prime}$ system.
Position vectors in $S$ are denoted by
${\bf x}=(x,y,z)$, or by ${\bf x}=(x_1,x_2,x_3)$ with
an obvious change in notation.
Writing time as a spatial coordinate $ct/n$, the
four-vector \cite{BIFinn1}
\begin{equation}
{\mathbb X}=(ct/n,{\bf x})=(x_0,x_1,x_2,x_3)
\label{EQb2.03}
\end{equation}
represents the position of a point in a filled Minkowski
spacetime.
In the modified Minkowski spacetime, the square of the invariant
spatial interval $\Delta s$ is \cite{BIFinn}
\begin{equation}
(\Delta s)^2=(\Delta x_1)^2+(\Delta x_2)^2+(\Delta x_3)^2-(\Delta x_0)^2
\label{EQb2.04}
\end{equation}
from which we obtain the interval of proper time
\begin{equation}
d\tau =\frac{dx_0}{\gamma c},
\label{EQb2.05}
\end{equation}
where
\begin{equation}
\gamma=\frac{1}{\sqrt{1-\frac{n^2v^2}{c^2}}}.
\label{EQb2.06}
\end{equation}
\par
Taking the derivative of the position four-vector (\ref{EQb2.03}) with
respect to the proper time, we obtain the
four-velocity \cite{BIquantclass}
\begin{equation}
{\mathbb U}=\frac{d{\mathbb X}}{d\tau}
=\frac{d{\mathbb X}}{dx_0}\frac{dx_0}{d\tau}
=\gamma c\left (1,\frac{dx_1}{dx_0},\frac{dx_2}{dx_0},
\frac{dx_3}{dx_0} \right )
\label{EQb2.07}
\end{equation}
and the four-momentum \cite{BIquantclass}
\begin{equation}
{\mathbb P}=m_0{\mathbb U}=
mc\left (1,\frac{dx_1}{dx_0},\frac{dx_2}{dx_0},
\frac{dx_3}{dx_0} \right ),
\label{EQb2.08}
\end{equation}
where $m=\gamma m_0$ is the relativistic mass.
In the nonrelativistic limit,
$\gamma=1$, $\tau=t/n$, the three-velocity
\begin{equation}
{\bf u}=n{\bf\dot x}=
c\left (\frac{dx_1}{dx_0},\frac{dx_2}{dx_0},\frac{dx_3}{dx_0}\right )
\label{EQb2.09}
\end{equation}
and three-momentum
\begin{equation}
{\bf p}=nm_0{\bf\dot x}=
m_0c\left (\frac{dx_1}{dx_0},\frac{dx_2}{dx_0},\frac{dx_3}{dx_0}\right )
\label{EQb2.10}
\end{equation}
are re-defined in a region of reduced light
velocity \cite{BIquantclass}.
\par
For a system of particles, the transformation of the position
vector ${\bf x}_i$ of the $i^{th}$ particle to $J$ independent
generalized coordinates is
\begin{equation}
{\bf x}_i={\bf x}_i(\tau;q_1,q_2, \ldots, q_J) ,
\label{EQb2.11}
\end{equation}
where $\tau=t/n$.
Applying the chain rule, we obtain the virtual displacement
\begin{equation}
\delta{\bf x}_i=\sum_{j=1}^J
\frac{\partial {\bf x}_i}{\partial q_j}\delta q_j
\label{EQb2.12}
\end{equation}
and the velocity
\begin{equation}
{\bf u}_i=\frac{d{\bf x}_i}{d\tau}=
\sum_{j=1}^J
\frac{\partial {\bf x}_i}{\partial q_j}
\frac{d q_j}{d \tau}
+ \frac{\partial {\bf x}_i}{\partial \tau}
\label{EQb2.13}
\end{equation}
of the $i^{th}$ particle in the new coordinate system.
Substitution of
\begin{equation}
\frac{\partial{\bf u}_i}{\partial(d q_j/d\tau)}=
\frac{\partial {\bf x}_i}{\partial q_j}
\label{EQb2.14}
\end{equation}
into the identity
\begin{equation}
\frac{d}{d\tau}\left ( m{\bf u}_i\cdot
\frac{\partial{\bf x}_i}{\partial q_j} \right ) =
m\frac{d{\bf u}_i}{d \tau}\cdot
\frac{\partial {\bf x}_i}{\partial q_j}
+
m{\bf u}_i\cdot\frac{d}{d\tau}
\left ( \frac{\partial{\bf x}_i}{\partial q_j}\right ) 
\label{EQb2.15}
\end{equation}
yields
\begin{equation}
\frac{d{\bf p}_i}{d\tau}\cdot
\frac{\partial{\bf x}_i}{\partial q_j} =
\frac{d}{d\tau}
\left ( \frac{\partial}{\partial(d q_j/d \tau)}
\frac{1}{2}m{\bf u}_i^2
\right ) -
\frac{\partial}{\partial q_j}\left ( \frac{1}{2}m{\bf u}_i^2\right ).
\label{EQb2.16}
\end{equation}
\par
For a system of particles in equilibrium, the virtual work of the
applied forces ${\bf f}_i$ vanishes and the virtual work on each
particle vanishes leading to the principle of virtual work
\begin{equation}
\sum_i{\bf f}_i\cdot \delta{\bf x}_i=0
\label{EQb2.17}
\end{equation}
and D'Alembert's principle
\begin{equation}
\sum_i\left ( {\bf f}_i -\frac{d{\bf p}_i}{d\tau}\right )
\cdot \delta{\bf x}_i=0.
\label{EQb2.18}
\end{equation}
Using Eqs.\ (\ref{EQb2.12}) and (\ref{EQb2.16})
and the kinetic energy of the $i^{th}$
particle 
\begin{equation}
T_i= \frac{1}{2} m{\bf u}_i^2,
\label{EQb2.19}
\end{equation}
we can write D'Alembert's principle, Eq.\ (\ref{EQb2.18}), as
\begin{equation}
\sum_j \left [ \left ( \frac{d}{d\tau}
\left (
\frac{\partial T}{\partial(d q_j/d \tau)}\right ) 
-\frac{\partial T}{\partial q_j}
\right ) -Q_j \right ] \delta q_j =0 ,
\label{EQb2.20}
\end{equation}
where
\begin{equation}
Q_j=\sum_i{\bf f}_i\cdot
\frac{\partial{\bf x}_i}{\partial q_j}.
\label{EQb2.21}
\end{equation}
If the generalized forces come from a generalized scalar
potential function $V$ \cite{BIGoldstein}, then we can write
Lagrange equations of motion
\begin{equation}
\frac{d}{d\tau} \left (
\frac{\partial L}{\partial(\partial q_j/\partial \tau)}\right ) 
- \frac{\partial L}{\partial q_j} =0,
\label{EQb2.22}
\end{equation}
where $L=T-V$ is the Lagrangian.
The canonical momentum is therefore \cite{BIquantclass}
\begin{equation}
p_j=\frac{\partial L}{\partial(d q_j/d \tau)}
=\frac{1}{c}\frac{\partial L}{\partial(d q_j/d x_0)}
\label{EQb2.23}
\end{equation}
in a linear medium.
This formula for the canonical momentum differs from the 
usual canonical momentum formula
\begin{equation}
p_j=\frac{\partial L}{\partial(d q_j/d t)}
\label{EQb2.24}
\end{equation}
by a factor of $n$.
\par
We can provide two examples that support the use of the new formulation
of the canonical momentum in a region of space in which light travels at
a speed of $c/n$.
First, the momentum formula, Eq.\ (\ref{EQb2.23}), applied to a free
particle traveling unimpeded in an arbitrarily large linear medium
yields
${\bf p}=nm{\bf \dot x}$ in
agreement with the result of special relativity,
Eq.\ (\ref{EQb2.10}).
Second, it was shown in Ref.\ \cite{BIquantclass} that application of
the new momentum formula (\ref{EQb2.23}) to the quantization of the
electromagnetic field in a dielectric repairs a violation of
quantum--classical correspondence caused by the vacuum formulation
(\ref{EQb2.24}) of the canonical momentum.
\par
The field theory \cite{BICT,BIHillMlod} is based on a generalization of
the discrete case in which the dynamics are derived from a Lagrangian
density ${\cal L}$ instead of the Lagrangian
\begin{equation}
L=\int dv {\cal L}.
\label{EQb2.25}
\end{equation}
The generalization of the Lagrange equation (\ref{EQb2.22})
for fields in a linear medium is
\begin{equation}
\frac{\partial}{\partial x_0}\frac{\partial{\cal L}}
{\partial (\partial A_{\nu}/\partial x_0)}
=\frac{\partial {\cal L}}{\partial A_{\nu}}
-\sum_i\partial_{i}
\frac{\partial{\cal L}}{\partial(\partial_{i} A_{\nu})},
\label{EQb2.26}
\end{equation}
where $x_0=ct/n$ is the time-like coordinate in the filled
Minkowski spacetime and $x_1$, $x_2$, and $x_3$ correspond to the
respective $x$, $y$ and $z$ coordinates.
We adopt the typical conventions that Roman indices run from one
to three, Greek indices run from zero to three, and $\partial_i$ 
represents the operator $\partial/\partial x_i$.
The conjugate momentum field
\begin{equation}
\Pi_{\nu}=
\frac{\partial{\cal L}}{\partial (\partial A_{\nu}/\partial x_0)}
\label{EQb2.27}
\end{equation}
is used to construct the Hamiltonian density
\begin{equation}
{\cal H}=
\sum_{\nu}\Pi_{\nu}\frac{\partial A_{\nu}}{\partial x_0}-{\cal L}
\label{EQb2.28}
\end{equation}
from which Hamilton's equations of motion
\begin{subequations}
\label{EQb2.29}
\begin{equation}
\frac{\partial A_{\nu}}{\partial x_0}
=
\frac{\partial{\cal H}}{\partial \Pi_{\nu}}
\label{EQb2.29a}
\end{equation}
\begin{equation}
\frac{\partial\Pi_{\nu}}{\partial x_0}=
-\frac{\partial{\cal H}}{\partial A_{\nu}}
+\sum_{i}\partial_{i}
\frac{\partial {\cal H}}{\partial(\partial_{i} A_{\nu})}
\label{EQb2.29b}
\end{equation}
\end{subequations}
are derived for an arbitrarily large region of space in which the
velocity of light is $c/n$.
\par
For the purpose of comparison, we quote the usual Hamilton's equations
of motion in free space,
\begin{subequations}
\label{EQb2.30}
\begin{equation}
\frac{1}{c}\frac{\partial A_{\nu}}{\partial t}
=
\frac{\partial{\cal H}}{\partial \Pi_{\nu}}
\label{EQb2.30a}
\end{equation}
\begin{equation}
\frac{1}{c}\frac{\partial\Pi_{\nu}}{\partial t}=
-\frac{\partial{\cal H}}{\partial A_{\nu}}
+\sum_{i}\partial_{i}
\frac{\partial {\cal H}}{\partial(\partial_{i} A_{\nu})},
\label{EQb2.30b}
\end{equation}
\end{subequations}
where
\begin{subequations}
\label{EQb2.31}
\begin{equation}
\Pi_{\nu}
=c\frac{\partial{\cal L}}{\partial(\partial A_{\nu}/\partial t)}
\label{EQb2.31a}
\end{equation}
\begin{equation}
{\cal H}=
\sum_{\nu}\frac{1}{c}\Pi_{\nu}\frac{\partial A_{\nu}}{\partial t}-{\cal L}.
\label{EQb2.31b}
\end{equation}
\end{subequations}
\par
\section{Hamilton's Equations for Fields}
\par
Before deriving the equations of motion of electromagnetic fields in
linear media, we review the
microscopic Maxwell equations so that they and their derivation may
serve as a basis for comparison with the present work.
The Lagrangian for the electromagnetic field in free space is typically
\begin{equation}
L=\int dv \frac{1}{2}
\left ( \frac{1}{c^2}\left ( 
\frac{\partial{\bf A}}{\partial t }\right )^2
- (\nabla\times{\bf A})^2 \right ).
\label{EQb3.01}
\end{equation}
Applying Eqs.\ (\ref{EQb2.31}) to the Lagrangian density, we obtain the
canonical momentum field in vacuum
\begin{equation}
{\bf \Pi} = \frac{1}{c}\frac{\partial {\bf A}}{\partial t}
\label{EQb3.02}
\end{equation}
and the Hamiltonian density
\begin{equation}
{\cal H}= \frac{1}{2}
\left (  {\bf \Pi}^2+(\nabla\times{\bf A})^2 \right ).
\label{EQb3.03}
\end{equation}
Then Hamilton's equations of motion, Eqs.\ (\ref{EQb2.30}), become
\begin{subequations}
\label{EQb3.04}
\begin{equation}
\frac{1}{c}\frac{\partial {\bf A}}{\partial t} = {\bf \Pi}
\label{EQb3.04a}
\end{equation}
\begin{equation}
\frac{1}{c}\frac{\partial{\bf \Pi}}{\partial t}
=-\nabla\times\nabla\times{\bf A}.
\label{EQb3.04b}
\end{equation}
\end{subequations}
\par
We define the magnetic field in terms of the vector
potential in the usual manner as
\begin{equation}
{\bf B}=\nabla\times {\bf A}.
\label{EQb3.05}
\end{equation}
Substituting  the definition of the magnetic field into
Eqs.\ (\ref{EQb3.04}), we obtain the microscopic Faraday and
Maxwell--Amp\`ere laws
\begin{subequations}
\label{EQb3.06}
\begin{equation}
\nabla\times {\bf \Pi} =\frac{\partial {\bf B}}{\partial x_0}
\label{EQb3.06a}
\end{equation}
\begin{equation}
\nabla\times{\bf B} =-\frac{\partial{\bf \Pi}}{\partial x_0}
\label{EQb3.06b}
\end{equation}
\end{subequations}
in terms of the canonical momentum field ${\bf \Pi}=-{\bf E}$
and timelike coordinate $x_0=ct$ of Minkowski spacetime.
\par
The filled Minkowski spacetime is the setting for the study of continuum
electrodynamics in which the linear medium acts like a region of space
in which the speed of light is $c/n$.
We take the Lagrangian of the electromagnetic field in the medium to be
\begin{equation}
L=
\int dv\frac{1}{2}
\left ( \left ( \frac{\partial{\bf A}}{\partial x_0 }\right )^2
- (\nabla\times{\bf A})^2 \right ),
\label{EQb3.07}
\end{equation}
where $x_0=ct/n$.
Applying Eq.\ (\ref{EQb2.27}), the canonical momentum field 
\begin{equation}
{\bf \Pi}= \frac{\partial {\bf A}}{\partial x_0}
\label{EQb3.08}
\end{equation}
is used to construct the Hamiltonian density
\begin{equation}
{\cal H}=\frac{1}{2}\left ({\bf \Pi}^2+(\nabla\times{\bf A})^2\right ).
\label{EQb3.09}
\end{equation}
Hamilton's equations of motion in the
filled spacetime
\begin{subequations}
\label{EQb3.10}
\begin{equation}
\frac{\partial {\bf A}}{\partial x_0}
=
{\bf \Pi}
\label{EQb3.10a}
\end{equation}
\begin{equation}
\frac{\partial{\bf \Pi}}{\partial x_0}
=-\nabla\times\nabla\times{\bf A}
\label{EQb3.10b}
\end{equation}
\end{subequations}
are obtained from the Hamiltonian density, Eq.\ (\ref{EQb3.09}),
using Eqs.\ (\ref{EQb2.29}).
\par
As an alternative, the universe can be modeled as a vacuum Minkowski
spacetime.
In that case, the equations of motion for electromagnetic fields in 
linear media are derived using Eqs.\ (\ref{EQb2.30}), instead of
Eqs.\ (\ref{EQb2.29}).
The Lagrangian of the electromagnetic field in a linear
medium of refractive index $n$, Eq.\ (\ref{EQb3.07}) can be written as
\begin{equation}
L=\int dv\frac{1}{2}
\left (
\frac{n^2}{c^2} \left (\frac{\partial{\bf A}}{\partial t }\right )^2
- {(\nabla\times{\bf A})^2} \right ).
\label{EQb3.11}
\end{equation}
In this case, the canonical momentum field
\begin{equation}
{\bf \Pi}_v= \frac{n^2}{c^2}\frac{\partial {\bf A}}{\partial t} 
\label{EQb3.12}
\end{equation}
is the same as the canonical momentum field that is used by
Hillery and Mlodinow \cite{BIHillMlod} and other authors.
The canonical momentum of Garrison and Chiao \cite{BIGarrChiao} is 
similar in name, but is a variant of the Minkowski electromagnetic
momentum.
In this case, the Hamiltonian density is
\begin{equation}
{\cal H}=\frac{1}{2}
\left (\frac{{\bf \Pi}_v^2}{n^2}
+ (\nabla\times{\bf A})^2 \right ),
\label{EQb3.13}
\end{equation}
from which Hamilton's equations of motion
\begin{subequations}
\label{EQb3.14}
\begin{equation}
\frac{1}{c}\frac{\partial{\bf A}}{\partial t}=\frac{c}{n^2}{\bf \Pi}_v
\label{EQb3.14a}
\end{equation}
\begin{equation}
\frac{1}{c}\frac{\partial{\bf \Pi}_v}{\partial t}
=-\nabla\times\nabla\times{\bf A}
\label{EQb3.14b}
\end{equation}
\end{subequations}
are derived using Eqs.\ (\ref{EQb2.30}).
\par
\section{Commutation Relations}
\par
Two different representations of equations of motion for the
macroscopic electromagnetic field in a linear medium have been
derived from the same Lagrangian.
In one case, the canonical momentum field was derived in a filled
Minkowski spacetime in which light travels at $c/n$.
In the other case, the canonical momentum field has its roots in the
special relativity of the vacuum.
Although the two representations do not differ much, we show that the
relativity of a filled Minkowski spacetime, rather than that of a vacuum
Minkowski spacetime, leads to satisfaction of quantum--classical
correspondence.
\par
The macroscopic field in a linear medium can be quantized in a manner
analogous to quantum electrodynamics.
In Ref.\ \cite{BIquantclass}, the macroscopic field was quantized in terms
of discrete modes of a quantization volume.
In that work, quantum--classical correspondence with respect to the
boundary conditions required a re-derivation of the discrete
D'Alembert's principle and the discrete Lagrange equations
of motion for dynamics in a filled Minkowski spacetime.
The current work extends the previous result to continuous fields.
\par
Quantization for continuous fields can be achieved by applying the
equal-time commutation relation \cite{BIHillMlod}
\begin{equation}
[{A}_i({\bf x},t),{\Pi}_j({\bf x}^{\prime},t)]
=i\delta_{ij}^{\rm tr}({\bf x}-{\bf x}^{\prime})
\label{EQb4.01}
\end{equation}
to the canonically conjugate fields.
For macroscopic fields, which satisfy the limit of large numbers, the 
boundary conditions on the quantized fields must be of the same form as
the classical electromagnetic boundary conditions in which the
amplitude of the vector potential inside the material is smaller than
the vacuum amplitude by $\sqrt{n}$, absent reflection.
\par
We have two candidates for the canonical momentum field in a linear
medium and the issue to be decided is which version satisfies both the
equal-time commutation relation and quantum--classical correspondence
with respect to the electromagnetic boundary conditions.
Relating the vector potential ${\bf A}$ inside the medium to the vector
potential ${\bf A}^0$ in the vacuum, the commutator
\begin{equation}
\left [ \frac{{A}^0_i({\bf x},t)}{\sqrt{n}}, \frac{c}{\sqrt{n}}
\frac{\partial {A}^0_j({\bf x}^{\prime},t)}{\partial (ct/n)} \right ]=
i\delta_{ij}^{\rm tr}({\bf x}-{\bf x}^{\prime})
\label{EQb4.02}
\end{equation}
is consistent with the equal-time commutation relation and boundary
conditions.
This commutator is based on the canonical momentum field (\ref{EQb3.08})
in a filled Minkoswski spacetime.
In contrast,  the commutator
\begin{equation}
\left [ \frac{{A}^0_i({\bf x},t)}{\sqrt{n}}, \frac{n^2}{\sqrt{n}}
\frac{\partial {A}^0_j({\bf x}^{\prime},t)}{\partial t} \right ]=
in\delta_{ij}^{\rm tr}({\bf x}-{\bf x}^{\prime})
\label{EQb4.03}
\end{equation}
for the canonical momentum field
(\ref{EQb3.12}) results in a contradiction between the commutation
relation and the boundary conditions, disproving Hamilton's
equations (\ref{EQb3.14}).
\par
The dynamical field theory of the filled Minkowski spacetime,
Eq.\ (\ref{EQb3.07})--Eq.\ (\ref{EQb3.10}), is consistent with the equal-time commutation
relation and the boundary conditions.
The substitution of the magnetic field ${\bf B}=\nabla\times{\bf A}$
into Hamilton's equations (\ref{EQb3.06}) produces equations of motion
\begin{subequations}
\label{EQb4.04}
\begin{equation}
\nabla\times{\bf \Pi}= \frac{\partial{\bf B}}{\partial x_0}
\label{EQb4.04a}
\end{equation}
\begin{equation}
\nabla\times{\bf B}=- \frac{\partial{\bf \Pi}}{\partial x_0}
\label{EQb4.04b}
\end{equation}
\end{subequations}
for the fields.
\par 
The field equations of motion that were derived relativistically in 
a filled Minkowski spacetime with timelike coordinate $x_0=ct/n$
and canonical momentum field ${\bf\Pi}=(n/c)\partial{\bf A}/\partial t$
are isomorphic with the the microscopic Faraday and Maxwell--Amp\`ere
laws derived in vacuum Minkowski spacetime.
On this basis, it is a trivial matter to transform from microscopic
electrodynamics in the vacuum to macroscopic electrodynamics in the
continuum limit of a linear medium.
Every occurrence of $ct$ that appears in free space electrodynamics
is replaced by $x_0$ and every occurrence of ${\bf E}$ is replaced by
$-{\bf \Pi}$.
Examples of electromagnetic quantities and equations that are valid both
microscopically and macroscopically are:
Poynting's theorem
\begin{equation}
\nabla\cdot ( {\bf B}\times{\bf \Pi})=
-\frac{\partial}{\partial x_0}
\frac{1}{2}\left ({\bf \Pi}^2+{\bf B}^2 \right ),
\label{EQb4.05}
\end{equation}
the electromagnetic energy
\begin{equation}
H=\int dv \frac{1}{2}\left ( {\bf \Pi}^2+{\bf B}^2 \right ) ,
\label{EQb4.06}
\end{equation}
the electromagnetic momentum
\begin{equation}
G=\int dv \frac{1}{c}\left ( {\bf B}\times{\bf \Pi} \right ) ,
\label{EQb4.07}
\end{equation}
the symmetric Maxwell stress tensor
\begin{equation}
T^{\alpha\beta}=\Pi_{\alpha}\Pi_{\beta} +B_{\alpha}B_{\beta} 
-\frac{1}{2}\left ( {\bf \Pi}\cdot{\bf \Pi}+ {\bf B}\cdot{\bf B}
\right ) \delta_{\alpha\beta},
\label{EQb4.08}
\end{equation}
the Lorentz force
\begin{equation}
{\bf F}=q\left (-{\bf \Pi}+\frac{d{\bf x}}{dx_0}\times{\bf B} \right ),
\label{EQb4.09}
\end{equation}
and the antisymmetric field tensor
\begin{equation}
F^{\alpha\beta}=
\left [
\begin{matrix}
 0        &-\Pi_x      &-\Pi_y      &-\Pi_y 
\cr
\Pi_x     &0           &-B_z        &B_y     
\cr
\Pi_y     &B_z         &0           &-B_x       
\cr
\Pi_z     &-B_y        &B_x         &0        
\cr
\end{matrix}
\right ].
\label{EQb4.10}
\end{equation}
\par
\par
\section{Superposition}
\par
For homogeneous media, the equations of motion for the fields,
Eqs.\ (\ref{EQb4.04}), can be transformed into the macroscopic
Faraday and Maxwell--Amp\`ere laws
\begin{subequations}
\label{EQb5.01}
\begin{equation}
\nabla\times{\bf E}= -\frac{1}{c}\frac{\partial{\bf B}}{\partial t}
\label{EQb5.01a}
\end{equation}
\begin{equation}
\nabla\times{\bf B}= \frac{n^2}{c}\frac{\partial{\bf E}}{\partial t}
\label{EQb5.01b}
\end{equation}
\end{subequations}
of classical continuum electrodynamics by defining an electric
field ${\bf E}=-{\bf \Pi}/n$.
However, one should be cautious about ascribing any
physical meaning to $n^2$ as anything other than the square of the
factor that scales the timelike coordinate of vacuum Minkowski spacetime
in a linear medium.
In particular, the disjunction
\begin{equation}
n^2=\varepsilon\mu
\label{EQb5.02}
\end{equation}
of the square of the refractive index into an electric permittivity
$\varepsilon$ and a magnetic permeability $\mu$ violates the principle
of superposition.
\par
Feynman \cite{BIFeynLect} defines the refractive index by the effect of
a slab of refractive material that is placed between the source of an
electromagnetic field and the point at which it is observed.
The macroscopic Feynman model can be easily extended to consider the
effect of a stack or mixture of refractive materials on the field.
Here, we derive the extended Feynman model and relate the result to the
microscopic model of Ward, Nelson, and Webb \cite{BIWNW} in which
the refractive slab of the Feynman model is replaced with thin sheets
of microscopic electric and magnetic dipoles.
\par
Consider a monochromatic source field of amplitude $E_0$ and frequency
$\omega$ that passes through a transparent plate of thickness $d$
and macroscopic refractive index $n$.
The electric field that is detected at a location $z$ is
\begin{equation}
E_d(z)=e^{-i \omega d(n-1) /c}E_0e^{i\omega(t-z/c)},
\label{EQb5.03}
\end{equation}
absent reflections that can be neglected if $\delta n=n-1$ is small
or if an anti-reflection coating is applied to the surfaces.
Expanding the first exponential in a power series, one obtains
\begin{equation}
E_d(z)=E_0e^{i\omega(t-z/c)}
-\frac{i\omega d(n-1)}{c}E_0 e^{i\omega(t-z/c)}
\label{EQb5.04}
\end{equation}
for small $d$ \cite{BIFeynLect}.
Now consider the slab to be composed of two layers of material of
thickness $d$ and refractive indices of $n_1$ and $n_2$.
The effect of the material on the detected field is
\begin{equation}
E_d(z)=e^{-i\omega d (\delta n_1+\delta n_2)/c}
E_0e^{i\omega(t-z/c)},
\label{EQb5.05}
\end{equation}
where an algebraic property of exponentials has been used to write the
product of exponentials as a single exponential by adding the exponents.
Again expanding the first exponential in a power series, produces
\begin{equation}
E_d(z)=E_0e^{i\omega(t-z/c)}
-\frac{i\omega(\delta n_1+\delta n_2 )d}{c}E_0 e^{i\omega(t-z/c)}.
\label{EQb5.06}
\end{equation}
Comparing Eqs.\ (\ref{EQb5.04}) and (\ref{EQb5.06}), one obtains
\begin{equation}
n=1+\delta n_1+\delta n_2
\label{EQb5.07}
\end{equation}
for the effective macroscopic index of refraction.
For small $\delta n$, we can make the approximation \cite{BIWNW}
\begin{equation}
n^2 \approx (1+\delta n_1)(1+\delta n_2).
\label{EQb5.08}
\end{equation}
For a material that responds linearly to electromagnetic radiation,
the refractive index must depend linearly on the material parameters.
Not only is Eq.\ (\ref{EQb5.08}) nonlinear, but there is no requirement
that the $\delta n$ are small because we have other techniques to
minimize reflections.
Extending the treatment to a stack of materials with layers indexed
with $i$, we obtain
\begin{equation}
n=1+\sum_i\delta n_i
\label{EQb5.09}
\end{equation}
as the definition of the effective macroscopic refractive index.
\par
Now consider the slab as a bounded region of space of refractive index
$n_1$ and thickness $d$ containing a density of particles.
As before,
\begin{equation}
E_d(z)=e^{-i \omega d(n_1-1) /c}E_0e^{i\omega(t-z/c)}.
\label{EQb5.10}
\end{equation}
If the volume emptied and then filled with a second group of particles
that create a refractive index of $n_2$, we get
\begin{equation}
E_d(z)=e^{-i \omega d(n_2-1) /c}E_0e^{i\omega(t-z/c)}.
\label{EQb5.11}
\end{equation}
Combining the two groups of particles in the same volume, we obtain
\begin{equation}
E_d(z)=e^{-i\omega d (\delta n_1+\delta n_2)/c}
E_0e^{i\omega(t-z/c)}.
\label{EQb5.12}
\end{equation}
by superposition.
Comparing Eqs.\ (\ref{EQb5.12}) and (\ref{EQb5.04}), one obtains
\begin{equation}
n=1+\delta n_1+\delta n_2
\label{EQb5.13}
\end{equation}
for the effective macroscopic index of refraction.
The approximation
\begin{equation}
n^2 \approx (1+\delta n_1)(1+\delta n_2)
\label{EQb5.14}
\end{equation}
is unnecessarily restrictive because there is no guarantee that the
$\delta n$ are small.
The approximation is clearly incorrect if the two groups of particles
are the same, since $n=1+2\delta n$ in that case.
\par
It should be noted that the materials that provided the refractive index
have not been specified.
The materials can be dielectric, magnetic, or magnetodielectric and in
any combination.
Then the effective refractive index obeys superposition and is the sum
of the contributions of the various materials.
Defining the susceptibility of each component as the contribution of
that component to the refractive index, the effective index
\begin{equation}
n=\sum_{i=0}\chi_i
\label{EQb5.15}
\end{equation}
is the linear combination of the susceptibilities of all the components
of the material, including the vacuum susceptibility $\chi_0=1$.
\par
\section{Boundary Conditions}
\par
Boundary conditions determine how fields in different homogeneous media
relate to each other.
We consider an electromagnetic field normally incident on an interface
between a half-space of Minkowski spacetime and a half-space of filled
Minkowski spacetime.
Propagating wave solutions of the wave equation can be represented by
the vector potential
\begin{equation}
{{\bf A}}_f= A_f \cos{(-(n\omega/c)x_0 +kz +\phi)}\hat {\bf e}_i
\label{EQb6.01}
\end{equation}
for forward traveling waves and by
\begin{equation}
{{\bf A}}_b= A_b \cos{(-(n\omega/c)x_0-kz +\phi)}\hat {\bf e}_i
\label{EQb6.02}
\end{equation}
for backward traveling waves.
Here, $\hat {\bf e}_i$ is a unit vector transverse to the direction
of propagation, ${k=n\omega/c}$, and $A_f$, $A_b$, $\phi$ are temporally
and spatially independent in the plane-wave cw regime.
The amplitudes of the vector potential for the incident, reflected, and
refracted fields are respectively denoted as $A_i$, $A_r$, and $A_t$.
\par
Conservation of energy provides a boundary condition and, for a 
flow, the conservation law is represented by a continuity equation for
the energy flux.
The Hamiltonian density (\ref{EQb3.09}) for the field in filled 
spacetime can be written as the energy density 
\begin{equation}
\rho_e=\frac{1}{2}\left ( {\bf \Pi}^2+{\bf B}^2  \right ) .
\label{EQb6.03}
\end{equation}
The Poynting--Umov vector ${\bf S}=\rho_e{\bf \dot x}$ is the continuous
energy flux vector that is associated with the energy conservation law.
In the plane-wave cw limit, 
\begin{equation}
A_i^2=A_r^2+nA_t^2
\label{EQb6.04}
\end{equation}
is obtained from continuity of the Poynting--Umov vector.
\par
A second continuity condition can be derived for the fields by applying
Stokes' theorem to a small loop that straddles the boundary.
The application of Stokes' theorem to Eq.\ (\ref{EQb4.04b})
$$
\nabla\times{\bf B}=- \frac{\partial{\bf \Pi}}{\partial x_0}
$$
yields a continuity equation for the magnetic field.
In terms of amplitudes, continuity of the magnetic field is expressed as
\begin{equation}
A_i+A_r=nA_t.
\label{EQb6.05}
\end{equation}
\par
Continuity of the magnetic field provides one factor of
Eq.\ (\ref{EQb6.04}).
The other factor 
\begin{equation}
A_i-A_r=A_t
\label{EQb6.06}
\end{equation}
can be interpreted as continuity of a vector potential where the
direction of the vector potential is reversed by reflection.
The linearly independent continuity equations (\ref{EQb6.05}) and
(\ref{EQb6.06}) can be combined algebraically to form the Fresnel
relations
\begin{subequations}
\label{EQb6.07}
\begin{equation}
\frac{A_r}{A_i}=\frac{n-1}{n+1}
\label{EQb6.07a}
\end{equation}
\begin{equation}
\frac{A_t}{A_i}=\frac{2}{n+1}.
\label{EQb6.07b}
\end{equation}
\end{subequations}
It is not particularly difficult to derive Fresnel-type boundary
conditions for oblique incidence or multiple materials \cite{BICMF}.
\par
There is no independent continuity equation that is associated with
the canonical momentum field, Eq.\ (\ref{EQb3.08}),
$$
{\bf \Pi}=\frac{\partial {\bf A}}{\partial x_0}.
$$
By the application of Stokes' law to Eq.\ (\ref{EQb4.04a}),
$$
\nabla\times{\bf \Pi}= \frac{\partial{\bf B}}{\partial x_0},
$$
we find that the continuity equations for the magnetic and momentum
fields are redundant as the conjugate momentum field changes sign upon
reflection.
There is one continuity equation associated with the conservation of
electromagnetic energy, Eq.\ (\ref{EQb6.06}), and one continuity
equation associated with the fields, Eq.\ (\ref{EQb6.03}).
In the case of the macroscopic Maxwell equations (\ref{EQb5.01}),
the boundary conditions are over-specified with two independent
continuity equations for the field and one continuity equation for the
energy flux.
\par
\section{Summary}
\par
We derived a theory of special relativity that applies to events that
occur in a dielectric or other linear medium that is treated as a
continuous material.
The relativistic theory was applied to the derivation of a
Lagrangian-based theory of particle and field dynamics in linear media
that was, in turn, used to develop equations of motion for macroscopic
fields.
We showed that these equations satisfy basic physical principles.
\par

\end{document}